\documentclass[twocolumn,showpacs,preprintnumbers,amsmath,amssymb,APSl,prd,nofootinbib,superscriptaddress]{revtex4-1}

\usepackage{dcolumn}% Align table columns on decimal point
\usepackage{bm}% bold math
\usepackage{ifpdf}
\usepackage{hyperref}
\usepackage{bm}
\usepackage{xcolor,color,graphicx,graphics}
\usepackage[spanish,english]{babel}%, portuguese
\usepackage[latin1]{inputenc}
\usepackage[OT1]{fontenc}
\usepackage{latexsym,amssymb,amsmath,amsfonts}
\usepackage{makeidx}
\usepackage{epsfig,subfigure}
\usepackage{natbib}
\usepackage{epstopdf}
\usepackage{mathrsfs}
\usepackage{hyperref}%\usepackage[colorlinks=true,linkcolor=blue]{hyperref}
\hypersetup{colorlinks=true, linkcolor=blue, citecolor=green}
\usepackage{enumerate}
\usepackage{tikz}

\definecolor{red}{rgb}{1,0,0}

\def\+{^\dagger}

\def\<{\leftarrow}
\def\>{\rightarrow}

\def\({\left(}
\def\){\right)}

\newcommand{\bi}{\begin{itemize}} 				\newcommand{\ei}{\end{itemize}}
\newcommand{\benu}{\begin{enumerate}} 		\newcommand{\enu}{\end{enumerate}}
\newcommand{\bd}{\begin{dinglist}{0}}     \newcommand{\ed}{\end{dinglist}}
\newcommand{\bfig}{\begin{figure}[htbp]}  \newcommand{\efig}{\end{figure}}
        			
\newcommand{\bc}{\begin{center}} 				  \newcommand{\ec}{\end{center}}
\newcommand{\be}{\begin{equation}} 				\newcommand{\ee}{\end{equation}}
\newcommand{\bsub}{\begin{subequations}}  \newcommand{\esub}{\end{subequations}}
\newcommand{\ben}{\begin{eqnarray}} 			\newcommand{\een}{\end{eqnarray}}
\newcommand{\ba}[1]{\begin{array}{#1}} 		\newcommand{\ea}{\end{array}}
\newcommand{\bea}{\begin{equation}\begin{array}{rcl}}
\newcommand{\eea}{\end{array}\end{equation}}

\definecolor{purple}{rgb}{1,0,1}
\definecolor{lime}{HTML}{A6CE39} % needs xcolor

\newcommand{\orcidicon}{%
	\begin{tikzpicture}
	\draw[lime, fill=lime] (0,0)
		circle [radius=0.16]
		node[white] {{\fontfamily{qag}\selectfont \tiny ID}};
	\draw[white, fill=white] (-0.0625,0.095)
		circle [radius=0.007];
	\end{tikzpicture}	\hspace{-2mm}
}
\newcommand\orcidGonzalo{{\href{https://orcid.org/0000-0001-9857-0412}{\orcidicon}}}
\newcommand\orcidDiego{{\href{https://orcid.org/0000-0003-3984-9864}{\orcidicon}}}
\newcommand\orcidFrancisco{{\href{https://orcid.org/0000-0002-9388-8373}{\orcidicon}}}

\begin{document}

%=================================================================

\title{Structure and stability of traversable thin-shell wormholes in Palatini $f(\mathcal{R})$ gravity}
%=================================================================

\author{Francisco S. N. Lobo\orcidFrancisco\!\!}
\email{fslobo@fc.ul.pt}
\affiliation{Instituto de Astrof\'{i}sica e Ci\^{e}ncias do Espa\c{c}o, Faculdade de Ci\^encias da Universidade de Lisboa, Edif\'{i}cio C8, Campo Grande, P-1749-016, Lisbon, Portugal}
%=================================================================
\author{Gonzalo J. Olmo\orcidGonzalo\!\!} \email{gonzalo.olmo@uv.es}
\affiliation{Departamento de F\'{i}sica Te\'{o}rica and IFIC, Centro Mixto Universidad de Valencia - CSIC.
Universidad de Valencia, Burjassot-46100, Valencia, Spain}
\affiliation{Departamento de F\'isica, Universidade Federal da
Para\'\i ba, 58051-900 Jo\~ao Pessoa, Para\'\i ba, Brazil}
%=================================================================
\author{Emanuele Orazi\orcidGonzalo\!\!} \email{orazi.emanuele@gmail.com}
\affiliation{ International Institute of Physics, Federal University of Rio Grande do Norte,
Campus Universit\'ario-Lagoa Nova, Natal-RN 59078-970, Brazil}
\affiliation{Escola de Ciencia e Tecnologia, Universidade Federal do Rio Grande do Norte, Caixa Postal 1524, Natal-RN 59078-970, Brazil}
%=================================================================
\author{Diego Rubiera-Garcia\orcidDiego\!\!} \email{drubiera@ucm.es}
\affiliation{Departamento de F\'isica Te\'orica and IPARCOS, Universidad Complutense de Madrid, E-28040 Madrid, Spain}
%=================================================================
\author{Azmat Rustam} \email{azmat.rustam@uv.es}
\affiliation{Departamento de F\'{i}sica Te\'{o}rica and IFIC, Centro Mixto Universidad de Valencia - CSIC.
Universidad de Valencia, Burjassot-46100, Valencia, Spain}
%=================================================================

\date{\today}
%%%%%%%%%%%%%%%%%%%%%%%%%%%%%%%%%%%%%%%%%%%%%%%%%%%%%%%%%%%%%%%%%
\begin{abstract}
%%%%%%%%%%%%%%%%%%%%%%%%%%%%%%%%%%%%%%%%%%%%%%%%%%%%%%%%%%%%%%%%%
We study the structure and stability of traversable wormholes built as (spherically symmetric) thin shells in the context of Palatini $f(\mathcal{R})$ gravity. Using a suitable junction formalism for these theories we find that the effective number of degrees of freedom on the shell is reduced to a single one, which fixes the equation
of state to be that of massless stress-energy fields, contrary to the general relativistic and metric $f(R)$ cases. Another major difference is that the surface energy density threading the thin-shell, needed in order to sustain the wormhole, can take any sign, and may even vanish, depending on the desired features of the corresponding solutions. We illustrate our results by constructing thin-shell wormholes by surgically grafting Schwarzschild space-times, and show that these configurations are always linearly unstable. However, surgically joined Reissner-Nordstr\"om space-times allow for linearly stable, traversable thin-shell wormholes supported by a positive energy density provided that the (squared) mass-to-charge ratio,  given by $y=Q^2/M^2$, satisfies the constraint $1<y<9/8$ (corresponding to overcharged Reissner-Nordstr\"om configurations having a photon sphere) and lies in a region bounded by  specific curves defined in terms of the (dimensionless) radius of the shell $x_0=R/M$.
%%%%%%%%%%%%%%%%%%%%%%%%%%%%%%%%%%%%%%%%%%%%%%%%%%%%%%%%%%%%%%%%%
\end{abstract}
%%%%%%%%%%%%%%%%%%%%%%%%%%%%%%%%%%%%%%%%%%%%%%%%%%%%%%%%%%%%%%%%%

\maketitle

%%%%%%%%%%%%%%%%%%%%%%%%%%%%%%%%%%%%%%%%%%%%%%%%%%%%%%%%%%%%%%%%%
\section{Introduction}
%%%%%%%%%%%%%%%%%%%%%%%%%%%%%%%%%%%%%%%%%%%%%%%%%%%%%%%%%%%%%%%%%

Wormholes are hypothetical objects connecting two widely separated regions of space-time or even two different universes. They arise as solutions of the Einstein field equations of General Relativity (GR) pretty much in the same way as black holes do. However, as opposed to the latter, for their physical plausibility wormholes lack two fundamental aspects: the fulfillment of the null energy condition \cite{MTT}, and the formulation of a well behaved collapse problem from initial conditions in physically realistic scenarios \cite{JoshiBook} (see however \cite{Deng:2016vzb}). The first issue is an unavoidable feature in GR, and in fact arises from the flaring-out condition, which is a fundamental ingredient in wormhole physics and follows from the requirement that any wormhole throat be sustainable against spontaneous collapse \cite{VisserBook,Lobo:2017oab}. However, it has been shown that this aspect can be relaxed in modified theories of gravity, where the matter threading the wormhole throat may satisfy all of the energy conditions, and it is the higher order curvature terms, which may be interpreted as a gravitational fluid,  that support these geometries \cite{Harko:2013yb}.

Among the  various solutions crafted in the literature are the so-called thin-shell wormholes \cite{Visser:1989kh,Visser:1989kg}, which are of special relevance for their applications \cite{Lemos:2003jb,Lobo:2004id,Lobo:2004rp,Lobo:2005zu,Lemos:2008aj,Eiroa:2004at,Rahaman:2006vg,Bronnikov:2009tv,Usmani:2010cd,Dias:2010uh,Mazharimousavi:2010bf,Halilsoy:2013iza,Eiroa:2015hrt,Eiroa:2016zjx,Shaikh:2018oul,Dzhunushaliev:2019qze,Lobo:2020jfl,Berry:2020tky}. These  geometries are obtained by surgically chopping and gluing together suitable patches of the same or different space-times at a certain hypersurface or {\it shell}  in such a way that no horizon is present, rendering a traversable wormhole. An advantage of this cut-and-paste construction is that the violations of the energy conditions are typically restricted to the shell, which can subsequently be made as small as possible to keep such violations restricted to a tiny region of the space-time. If we focus our attention  to asymptotically flat, spherically symmetric space-times, the simplest case is that of joining two vacuum configurations, such as the Schwarzschild \cite{Poisson:1995sv} and the Reissner-Nordstr\"om configurations \cite{Eiroa:2003wp}, though generalizations to include the presence of a cosmological constant \cite{Lobo:2003xd} or other constructions are certainly possible \cite{Garcia:2011aa,Lobo:2020kxn}. Since at the matching hypersurface discontinuities in several geometrical quantities can occur, this problem requires an upgrade of the concept of tensorial functions to that of tensorial distributions to rigorously deal with it. The net result is that a number of conditions on several geometrical and matter quantities must hold at the shell and across it, together with some additional relations from the energy conservation equations. These are known as the Darmois-Israel matching or {\it junction conditions} \cite{Israel:1966rt,Darmois,Mars:1993mj,Musgrave:1995ka}.

In a recent paper, some of us \cite{OlRuJC} have found the junction conditions of a simple and natural extension of GR, namely, $f(\mathcal{R})$ theories of gravity.  As opposed to the analysis of \cite{Senovilla:2013vra}, where $f(R)$ theories in the metric formalism were considered (with the affine connection set to be given {\it ab initio} by the Christoffel symbols of the metric), in \cite{OlRuJC} the Palatini (or metric-affine) formulation\footnote{From now on we will denote the Palatini Ricci scalar by $\mathcal{R}$, and the more conventional metric one by $R$ to avoid confusions}, where metric and affine connection are regarded as independent entities \cite{Olmo:2011uz}, was discussed. Restoring their freedom to these two entities has a large impact on the properties and dynamics of the corresponding solutions of such theories, and indeed a large body of results has been released in the last few years in black hole physics and cosmology, see e.g. \cite{Bambi:2015zch,Antoniadis:2018ywb,Olmo:2019qsj,Gialamas:2019nly,Rubio:2019ypq,Wojnar:2020txr}. The corresponding junction conditions in the Palatini formalism depart from those of the metric one, and only coincide when the action is given by the Einstein-Hilbert one (with a cosmological constant term), where both formulations recover GR solutions and junction conditions. This fact has a non-trivial impact in the implementation of specific applications where junction conditions are needed, such as in stellar models, domain walls, braneworld models, etc.

The main aim of the present work is to apply these junction conditions to build spherically symmetric thin-shell wormholes. We shall show that  their implementation will reduce the corresponding effective degrees of freedom on the shell from two (as in GR or metric $f(R)$ gravity) down to one, which in turns fixes the equation of state to be that of massless stress-energy fields, something not possible within GR. As a specific example we apply this result to build traversable wormholes from surgically joined Schwarzschild and Reissner-Nordstr\"om space-times, respectively. We shall show that in both these cases the energy density needed to sustain the wormhole throat can take any sign (and even vanish!), depending on the desired features of the corresponding solutions and, therefore, the need for exotic sources of matter can be avoided within these theories. However, in the Schwarzschild case there is a growing mode under linear perturbations for any $f(\mathcal{R})$ Lagrangian rendering the corresponding thin-shell wormhole unstable. As opposed to that result, we find that in the Reissner-Nordstr\"om case (linearly) stable traversable thin-shell wormholes supported by matter-energy sources with finite and positive energy density are possible provided that the (squared) charge-to-mass ratio, given by $y=Q^2/M^2$, satisfies $1<y <9/8$, and that the dimensionless radius of the matching hypersurface,  $x_0=R/M$,  is constrained to a region limited by certain curves  $y_-(x_0)<y<\frac{x_0}{2}(3-x_0)$ (see below for more details). This yields a new family of horizonless compact objects having a photon sphere (home to their innermost circular unstable orbits), with important potential applications.

This work is organized as follows:  In Sec. \ref{sec:II}, we introduce the junction conditions for Palatini $f(\mathcal{R})$ gravity and particularize them to the case of spherically symmetric space-times. In Sec. \ref{sec:III}, we use this formalism to build thin-shell wormholes from surgically joined Schwarzschild and Reissner-Nordstr\"om space-times and discuss the stability properties and positivity of the energy density of each case. A numerical analysis on the stable configurations of the Reissner-Nordstr\"om case is carried out in Sec. \ref{sec:IV} to complement these results. Finally, Sec. \ref{sec:V} contains our conclusion and a discussion of the obtained results and future perspectives.

%%%%%%%%%%%%%%%%%%%%%%%%%%%%%%%%%%%%%%%%%%%%%%%%%%%%%%%%%%%%%%%%%
\section{Junction formalism for Palatini $f(\mathcal{R})$ gravity} \label{sec:II}
%%%%%%%%%%%%%%%%%%%%%%%%%%%%%%%%%%%%%%%%%%%%%%%%%%%%%%%%%%%%%%%%%

%%%%%%%%%%%%%%%%%%%%%%%%%%%%%%%%%%%%%%%%%%%%%%%%%%%%%%%%%%%%%%%%%
\subsection{General form of the junction conditions}
%%%%%%%%%%%%%%%%%%%%%%%%%%%%%%%%%%%%%%%%%%%%%%%%%%%%%%%%%%%%%%%%%

Let us consider two smooth manifolds $\mathcal{M}_{\pm}$ with boundaries $\partial \mathcal{M}_{\pm}\equiv\Sigma_{\pm}$ which are assumed to be isometric, that is, $\Sigma_+=\Sigma_-$. Let us join these two bulks to define a new manifold $\mathcal{M} = \mathcal{M}_+ \cap \mathcal{M}_-$ with no boundary but containing a time-like hypersurface $\Sigma$ where $\mathcal{M}_{+}$ and $\mathcal{M}_{-}$ are pasted at,  and across which the space-time metric components of $g_{\mu\nu}$ must be continuous (but not necessarily their derivatives). In order to find the conditions that the geometrical quantities need to fulfill for this construction to be well defined from a mathematical point of view, one needs to call upon the theory of tensorial distributions. The specific form of these conditions depends on the theory of gravity under consideration, so we shall start our analysis from this point.

The action of $f(\mathcal{R})$ gravity is given by
\begin{equation} \label{eq:actionfR}
\mathcal{S}=\frac{1}{2\kappa^2} \int d^4x \sqrt{-g}  f(\mathcal{R}) + \mathcal{S}_m(g_{\mu\nu},\psi_m) \ ,
\end{equation}
where $\kappa^2$ is Newton's constant in suitable units, $g$ is the determinant of the space-time metric $g_{\mu\nu}$, while $f(\mathcal{R})$ is some function of the curvature scalar $\mathcal{R} \equiv g^{\mu\nu}\mathcal{R}_{\mu\nu}$,  defined from the Ricci tensor as $\mathcal{R}_{\mu\nu}(\Gamma) \equiv {\mathcal{R}^\alpha}_{\alpha\mu\nu}(\Gamma)$. Note that, since we are working in the Palatini formalism, metric and affine connection are independent entities, which implies that the Ricci tensor is determined solely by the affine connection $\Gamma \equiv \Gamma_{\mu\nu}^{\lambda}$. As for the matter action, $\mathcal{S}_m=\int d^4 x \sqrt{-g} \mathcal{L}_m(g_{\mu\nu},\psi_m)$, it is assumed to depend only on the space-time metric and on the matter fields $\psi_m$, but not on the connection, which can be done since in this paper we are dealing with bosonic fields alone.

The trademark of this theory and formulation is the particular way the matter fields source the new gravitational dynamics. Indeed, the trace of the equation associated to the variation of the action (\ref{eq:actionfR}) reads
\begin{equation} \label{eq:traceq}
\mathcal{R}f_\mathcal{R}-2f=\kappa^2 T \ ,
\end{equation}
where $f_\mathcal{R} \equiv df/d\mathcal{R}$ while $T \equiv g^{\mu\nu}T_{\mu\nu}$ is the trace of the stress-energy tensor of the matter fields, $  T_{\mu\nu}= \frac{2}{\sqrt{-g}} \frac{\delta \mathcal{S}_m}{\delta g^{\mu\nu}}$. This means that, as opposed to their metric $f(R)$ counterparts, where this equation includes derivative operators \cite{Olmo:2006eh,Olmo:2005zr,Olmo:2005hc}, therefore implying the propagation of an extra scalar degree of freedom, in the Palatini case it is instead an algebraic equation that can be solved as $\mathcal{R} \equiv \mathcal{R}(T)$ \cite{Olmo:2011uz}. This result allows to remove the curvature scalar in favour of the matter fields, with the side effect that no other propagating degrees of freedom than the two polarizations of the gravitational field travelling at the speed of light are introduced.  In vacuum, ${T^\mu}_{\nu}=0$, or for a traceless stress-energy tensor, $T=0$, this theory recovers GR dynamics and solutions, possibly with a cosmological constant term.

The key role played by the trace of the stress-energy tensor in the dynamics of this theory also manifests itself in the associated junction conditions. Indeed, denoting by ${S^\mu}_{\nu}$ the singular part of the stress-energy tensor on $\Sigma$, with $S \equiv {S^\mu}_{\mu}$ its trace, it has been obtained in Ref.\cite{OlRuJC} that the following junction conditions must hold:
\begin{equation} \label{eq:jc}
[T]=0 \,, \quad  \quad S=0 \ ,
\end{equation}
where from now on brackets will denote discontinuities (``jumps") across the hypersurface $\Sigma$ of the quantity inside them, for instance, $[T]=T^+\vert_{\Sigma} -T^-\vert_{\Sigma}$ is the discontinuity of the trace of the stress-energy tensor across $\Sigma$. Therefore, these junction conditions  (\ref{eq:jc}) simply tell us that both this discontinuity and the trace of the stress-energy tensor on $\Sigma$ must vanish. Note at this point that in GR the first such condition is not present, since this comes from a new contribution in the field equations of Palatini $f(\mathcal{R})$ gravity.

There are further junction conditions describing the interaction between matter and gravity on the matching surface. By defining the pull-back of the first fundamental form to $\mathcal{M}_{\pm}$ (that is, the projector or induced metric on $\Sigma$) as $h_{\mu\nu}=g_{\mu\nu}-n_{\mu}n_{\nu}$, where $n_{\mu}$ is the unit vector normal to $\Sigma$, and the pull-back of the second fundamental form (the extrinsic curvature) on both sides of $\mathcal{M}^{\pm}$ as
\begin{equation} \label{eq:sff}
K_{\mu\nu}^{\pm} \equiv {h^\rho}_{\mu} {h^\sigma}_{\nu} \nabla_{\rho}^{\pm} n_{\sigma} \ ,
\end{equation}
from the singular part of the field equations one finds the relation  \cite{OlRuJC}
\begin{equation} \label{eq:Kfinal}
-[K_{\mu\nu}]+\frac{1}{3}h_{\mu\nu}[K^{\rho}_{\rho}]=\kappa^2 \frac{S_{\mu\nu}}{f_{\mathcal{R}_\Sigma}} \ ,
\end{equation}
where the subindex $_{\Sigma}$  denotes quantities evaluated on $\Sigma$, and indices are raised with $h^{\mu\nu}$. This equation is consistent with the traceless character of $S_{\mu\nu}$ obtained in (\ref{eq:jc}), and  departs from the GR one, which reads $-[K_{\mu\nu}]+h_{\mu\nu}[K^{\rho}_{\rho}]=\kappa^2 S_{\mu\nu}$, and moreover it does not recover it smoothly in the limit $f_\mathcal{R} \to 1$. The reason is that in Palatini $f(\mathcal{R})$ gravity, due to the second constraint of the junction conditions (\ref{eq:jc}), the singular part of the Einstein equations takes one to the result $[{K^\rho}_{\rho}]=-\frac{3f_{RR}R_T b}{2f_R}$, where $b \equiv n^{\mu}[\nabla_{\mu} T]$ and $f_{\mathcal{R}\mathcal{R}} \equiv d^2 f/d\mathcal{R}^2$. This equation is non-vanishing as long as $f(\mathcal{R}) \neq \mathcal{R}-2\Lambda$ (the Einstein-Hilbert action of GR with a cosmological constant term). In the latter case one finds instead $[{K^\rho}_{\rho}]=S/2$, which is non-vanishing due to the fact that in GR the quantity $S$ is unconstrained.

On the other hand, the conservation of energy implies that
\begin{equation} \label{eq:main2}
D^{\rho}S_{\rho\nu}=-n^{\rho}{h^\sigma}_{\nu} [T_{\rho\sigma}] \ ,
\end{equation}
which relates the energy content on the shell with the discontinuity in the stress-energy tensor across it (since only its trace is constrained by (\ref{eq:jc})). In addition, from the Bianchi identities one also finds that
\begin{equation} \label{eq:main3}
(K_{\rho\sigma}^+ + K_{\rho\sigma}^-)S^{\rho\sigma} =2n^{\rho}n^{\sigma}[T_{\rho\sigma}] -\frac{3\mathcal{R}_T^2 f_{\mathcal{R}\mathcal{R}}^2}{f_\mathcal{R}} [b^2] \ .
\end{equation}
where $\mathcal{R}_T \equiv d\mathcal{R}/dT$. Therefore, the junction conditions for Palatini $f(\mathcal{R})$ gravity are those on the matter fields (\ref{eq:jc}) across the shell, and the relations between gravity and the matter given by Eqs.(\ref{eq:Kfinal}), (\ref{eq:main2}) and (\ref{eq:main3}).

%%%%%%%%%%%%%%%%%%%%%%%%%%%%%%%%%%%%%%%%%%%%%%%%%%%%%%%%%%%%%%%%%
\subsection{Junction conditions for spherically symmetric space-times}
%%%%%%%%%%%%%%%%%%%%%%%%%%%%%%%%%%%%%%%%%%%%%%%%%%%%%%%%%%%%%%%%%

Let us now use the above junction conditions to formulate the corresponding equations for general spherically symmetric space-times. In the hypersurface $\Sigma$ we have three independent basis vectors $e_i\equiv \partial/\partial \xi^i$ with components $e^\mu_i=\partial x^\mu/\partial \xi^i$, being  $\xi^i$ the coordinates on $\Sigma$. The induced metric on $\Sigma$  is then expressed as $h_{ij}=g_{\mu\nu}e^\mu_i e^\nu_j$, and we note that $n_\mu e^\mu_i=0$. The second fundamental form (\ref{eq:sfff})  is thus defined by $K_{ij}=e^\mu_i e^\nu_j \nabla_\mu n_\nu$, which is symmetric. Now, differentiating  $n_\mu e^\mu_i=0$ with respect to $\xi^j$, one can write the useful formula
\begin{equation} \label{eq:sfff}
K^{\pm}_{ij}= -n_\mu\left(\frac{\partial^2 x^\mu}{\partial \xi^i \partial \xi^j }+\Gamma^{\mu\pm}_{\alpha\beta}\frac{\partial x^\alpha}{\partial\xi^i}\frac{\partial x^\beta}{\partial\xi^j}\right) \ ,
\end{equation}
for the computation of the second fundamental form. Let us now introduce two spherically symmetric space-times whose line elements on $\mathcal{M}_{\pm}$ are
\begin{equation} \label{eq:SSSfull}
ds^2_\pm=-A_\pm(r_{\pm}) dt^2+\frac{1}{B_\pm(r_{\pm}) }dr^2_\pm +r^2_\pm d\Omega^2 \ ,
\end{equation}
described by the functions $A_{\pm}(r_{\pm}),B_{\pm}(r_{\pm})$, respectively, and where $d\Omega^2=d\theta^2 + \sin^2 \theta d\varphi^2$ is the unit volume in the two-spheres.

The induced line element on the matching hypersurface $\Sigma$ can be written as
\begin{equation}
ds^2_\Sigma=-d\tau^2+R^2(\tau) d\Omega^2 \ ,
\end{equation}
which is parameterized in terms of the proper time of an observer comoving with $\Sigma$. Here $R$ is the radius of the shell and $4\pi R^2(\tau)$ measures its area. This spherical three-dimensional hypersurface has coordinates $x^\mu(\tau,\theta,\varphi)=(t(\tau),R(\tau),\theta,\varphi)$, and since the tangent vectors to it are $e_\theta^\mu=(0,0,1,0)$ and $e_\varphi^\mu=(0,0,0,1)$, setting the velocity vector as $U^\mu \equiv dx^{\mu}/d\tau=(t_\tau, R_\tau,0,0)$ (where $t_{\tau} \equiv dt/d\tau, R_{\tau} \equiv dR/d\tau$), it follows that the normal vector (assumed to be oriented from $\mathcal{M}_-$ to $\mathcal{M}_+$) must be of the form $n^\mu=\pm(n^t,n^r,0,0)$. Now, using the fact that $n^\mu n_\nu=+1$ (space-like character) and $n_\mu U^\mu=0$ (orthogonality condition) one finds that the components of the normal vector, $n^t$ and $n^r$,  are given by
\begin{eqnarray}
n^t&=& \frac{R_\tau}{\sqrt{AB}}\,, \\
n^r&=& \sqrt{B+R_\tau^2} \ ,
\end{eqnarray}
where the metric functions $A$ and $B$ are the evaluation of the functions of (\ref{eq:SSSfull}) on $\mathcal{M}_{\pm}$  at the shell radius $r=R(\tau)$ and which, due to the condition on the continuity of the space-time metric across $\Sigma$, must match there.  With these formulae, it is now immediate to compute the non-vanishing components of the second fundamental form ${K^i}_{j}=\text{diag}({K^\tau}_{\tau},{K^\theta}_{\theta},{K^\theta}_{\theta})$, which from (\ref{eq:sfff}) and (\ref{eq:SSSfull}) turn out to be
\begin{eqnarray}
{K^\tau_{\ \tau}}^{\pm}&=&\pm \frac{B^2 A_R+(BA_R-A B_R) R_\tau^2+2 AB R_{\tau \tau}}{2A B\sqrt{B+R_\tau^2}}, \label{eq:Ktt}\\
{K^\theta_{\ \theta}}^{\pm}&=&\pm\frac{\sqrt{B+R_\tau^2}}{R} \label{eq:Kthethe}  \ .
\end{eqnarray}
where $A_R\equiv dA/dR$ and so on.

The matter stress-energy tensor on $\Sigma$ can also be written as a diagonal matrix ${S^\mu}_{\nu}=\text{diag}({S^\tau}_{\tau},{S^\theta}_{\theta},{S^\theta}_{\theta})$ with components $S^\tau_{\ \tau}=-\sigma$ and $S^\theta_{\ \theta}=\mathcal{P}$, where $\sigma$ is the surface energy density and $\mathcal{P}$ is the tangential surface pressure. Recall that via the second condition of Eq.(\ref{eq:jc}) this is subject to the constraint $S=-\sigma +2\mathcal{P}=0$, which implies that the pressure $\mathcal{P}=\sigma/2$ is completely determined by the energy density $\sigma$, thus reducing the effective $f(\mathcal{R})$ number of degrees of freedom just to one. This is a strong departure from the results in both GR and in the metric formulation of $f(\mathcal{R})$ gravity, where one typically needs an equation of state of the form $\mathcal{P}=\mathcal{P}(\sigma)$ to close the system. This condition actually describes a massless stress-energy source confined to the hypersurface, and it is commonly found in the Casimir effect involving massless fields \cite{MTY, Visser:1989kh}. It is worth pointing out that in GR it is not possible to find any solution of the Einstein field equations for this matter source, since it would render both $\sigma$ and $\mathcal{P}$ complex. As opposed to that no-go result, here we have no constraints forbidding us to build thin-shell wormholes upon  these conditions, as we shall see below.

Under this setting, the field equations on the shell (\ref{eq:Kfinal}) can thus be combined into a single equation of the form
\begin{equation}\label{eq:EOM1}
[K^\tau_{\ \tau}]- [K^\theta_{\ \theta}]=\frac{3\kappa^2}{2f_{R_\Sigma}}\sigma \ ,
\end{equation}
which reduces the problem of computing the energy density of the system to the substraction of the discontinuity of the second fundamental forms. Note that, since the trace of $T_{\mu\nu}$ in the bulk must be continuous, thanks to (\ref{eq:traceq}), then $f_{R_\Sigma}$ is also a continuous function evaluated at $\Sigma$ (because $f(R) \equiv f(R(T))$).
We finally need to consider the energy conservation equation (\ref{eq:main2}), which in the spherically symmetric case studied here implies that
\begin{equation}
-D_\rho S^{\rho}_{\ \nu}=\dot\sigma+\frac{2\dot R}{R}(\sigma+\mathcal{P}) \ .
\end{equation}
Taking into account that $\mathcal{P}=\sigma/2$, this quantity can be written as
\begin{equation} \label{eq:flux}
-D_\rho S^{\rho}_{\ \nu}=\frac{1}{R^3}\frac{d(\sigma R^3)}{d\tau}\ ,
\end{equation}
which via the junction condition (\ref{eq:main2}) allows us to find specific solutions once the discontinuity in the stress-energy tensor across the shell is set. In the simplest possible scenario, where $n^{\rho}{h^\sigma}_{\nu} [T_{\rho\sigma}]$ vanishes, this equations admits the simple solution (with $C$ an integration constant)
\begin{equation} \label{eq:sigmacon}
\sigma=\frac{C}{R^3} \ ,
\end{equation}
which relates the energy density with the radius of the shell. This is the case, in particular, of the vacuum and electrovacuum scenarios we will be considering in this paper. Since in this scenario the pressure inherits the sign of $\sigma$ (recall that $\mathcal{P}=\sigma/2$), this means that the positivity of the constant $C$ entails the fulfillment of all pointwise energy conditions. This concludes our analysis of the junction conditions for general spherically symmetric space-times.  In the next section we shall study specific examples of this setting.

%%%%%%%%%%%%%%%%%%%%%%%%%%%%%%%%%%%%%%%%%%%%%%%%%%%%%%%%%%%%%%%%%
\section{Traversable  wormholes from surgically joined space-times} \label{sec:III}
%%%%%%%%%%%%%%%%%%%%%%%%%%%%%%%%%%%%%%%%%%%%%%%%%%%%%%%%%%%%%%%%%

In this section we will proceed to study the existence and stability of thin-shell wormholes which are constructed by matching two identical spherically symmetric space-times at a given hypersurface. In particular, we will join two Schwarzschild space-times and two Reissner-Nordstr\"om space-times, respectively.

%%%%%%%%%%%%%%%%%%%%%%%%%%%%%%%%%%%%%%%%%%%%%%%%%%%%%%%%%%%%%%%%%
\subsection{Schwarzschild space-times}
%%%%%%%%%%%%%%%%%%%%%%%%%%%%%%%%%%%%%%%%%%%%%%%%%%%%%%%%%%%%%%%%%

Following Visser \cite{VisserBook,Visser:1989kg,Poisson:1995sv}, we consider a traversable wormhole constructed by joining two exterior Schwarzschild solutions at a spherical hypersurface of radius $R=R_0>2M$ (the throat) to avoid the presence (on both sides) of a horizon. This type of surgically  grafted space-time is particularly interesting because it restores geodesic completeness and implies a vanishing stress-energy tensor everywhere except at the matching hypersurface and, therefore, the energy density in such a case is described by (\ref{eq:sigmacon}). As for the throat of the wormhole, located at $\Sigma$, it can be viewed as a (planar) domain wall interpolating between the two regions $\mathcal{M}_{\pm}$.
On the other hand, the field equations (\ref{eq:EOM1}) will provide another relation between $R(\tau)$ and $\sigma(R)$, which will allow us to find a complete solution to the problem. Note that since $T_{\mu\nu}$ vanishes in $\mathcal{M}_\pm$,  by the first condition of Eq.(\ref{eq:jc}) we must have that $T=0$ everywhere, even on $\Sigma$. This means that, regardless of the form of the (nonlinear) Lagrangian $f(\mathcal{R})$, we must have $\left.\kappa^2/f_{R_\Sigma}\right|_{T=0}=\tilde{\kappa}^2={\rm constant}$, which is a remarkable universal property.  Given that for the Schwarzschild geometry we have $A=B^{-1}=1-2M/r$, then we find that $[K^\mu_{\ \nu}]=2{K^\mu_{\ \nu}}^+$, while Eq.(\ref{eq:main3}) vanishes identically.

By inserting the expressions (\ref{eq:Ktt}) and (\ref{eq:Kthethe}) into the left-hand side of Eq.(\ref{eq:EOM1}) and taking into account the discussion above, we get
\begin{equation} \label{eq:KKeq}
\frac{3M-R(1+R_\tau^2)+R^2 R_{\tau\tau}}{3R^2\sqrt{1-\frac{2M}{R}+R_\tau^2}}=\frac{\tilde{\kappa}^2}{4}\sigma=\frac{\tilde{\kappa}^2C}{4R^3} \ ,
\end{equation}
which can be manipulated to obtain
\begin{equation} \label{eq:Rtautau}
R_{\tau\tau}=\frac{R_\tau^2}{R}+\frac{1}{R}\left(1-\frac{3M}{R}\right)+\frac{\gamma}{R^3} {\sqrt{1-\frac{2M}{R}+R_\tau^2}}\ ,
\end{equation}
where we have introduced the constant $\gamma \equiv \frac{3\tilde{\kappa}^2 C}{4}$. This equation is particularly useful to study static configurations and their perturbations.  In fact, assuming that there is an equilibrium point with $R_{\tau}=0$ at the shell, $R_0>2M$, we can expand the right-hand side of (\ref{eq:Rtautau}) in a series of powers of $(R-R_0)$ to explore the conditions which lead to small, stable (oscillatory) perturbations about this equilibrium point \cite{Ishak:2001az,Eiroa:2007qz,Yue:2011cq}. This way we find
\begin{eqnarray} \label{eq:RHSeq}
R_{\tau\tau} &\approx& \left(\frac{R_0-3M}{R_0^2}+\gamma \frac{\sqrt{R_0-2M}}{R_0^{7/2}} \right) \\
&&+\left(\frac{6M-R_0}{R_0^3}-\gamma \frac{3R_0-7M}{R_0^{9/2} \sqrt{R_0-2M} }\right)(R-R_0)   \nonumber  \\
&&+ \mathcal{O}(R-R_0)^2 \,.  \nonumber
\end{eqnarray}

Any equilibrium points can be found by determining the values of $\gamma$ for which  the zeroth-order term in this expansion vanishes, and such that the resulting equation for the perturbations has bounded amplitude. Taking into account the expansion above, we have
\begin{equation} \label{eq:gammae}
\gamma_e=\frac{R_0^{3/2}(3M-R_0)}{\sqrt{R_0-2M}} \ ,
\end{equation}
which indicates that if $2M<R_0\leq 3M$ equilibrium configurations supported by positive energy density are possible (recall Eq.(\ref{eq:sigmacon}) and that $\gamma \equiv \frac{3\tilde{\kappa}^2 C}{4}$). Inserting this result back in Eq.(\ref{eq:RHSeq}) we get (at linear order in the expansion)
\begin{equation} \label{eq:deltasch}
\delta_{\tau \tau}+\omega_{Sch}^2 \delta(\tau)=0 \ ,
\end{equation}
where $\delta \equiv R-R_0$ and
\begin{equation}
\omega_{Sch}^2= \frac{-2 R_0^2+8MR_0-9M^2}{(R_0-2M) R_0^3 } \ .
\end{equation}
As can be easily verified, for a matching surface $\Sigma$ beyond the Schwarzschild radius, $R_0>2M$, this coefficient  is always negative, thus yielding a growing mode. This means that pure thin-shell Schwarzschild wormholes in Palatini $f(\mathcal{R})$ gravity are always unstable under linear perturbations regardless of the form of the $f(\mathcal{R})$ Lagrangian chosen.

%%%%%%%%%%%%%%%%%%%%%%%%%%%%%%%%%%%%%%%%%%%%%%%%%%%%%%%%%%%%%%%%%
\subsection{Reissner-Nordstr\"om space-times}
%%%%%%%%%%%%%%%%%%%%%%%%%%%%%%%%%%%%%%%%%%%%%%%%%%%%%%%%%%%%%%%%%

Let us now consider  two Reissner-Nordstr\"om geometries cut and pasted at a spherical hypersurface of radius $R$. Since we have again $n^{\rho}{h^\sigma}_{\nu} [T_{\rho\sigma}]=0$, then Eq.(\ref{eq:sigmacon}) still holds. Moreover, since for a Maxwell field $T=0$, then it follows again that $\mathcal{R} ={\rm constant}$ and that both of the junction conditions (\ref{eq:jc}) automatically hold, which implies that we can use pretty much the same expressions as in the Schwarzschild case above. Indeed, setting $A=B^{-1}=1-2M/r+Q^2/r^2$, one finds that (\ref{eq:KKeq}) generalizes to
\begin{equation}
\frac{3MR-2Q^2-R^2(1+R_{\tau}^2)+R^3R_{\tau \tau}}{3R^3 \sqrt{1-\frac{2M}{R}+\frac{Q^2}{R^2} + R_{\tau}^2}}=\frac{\tilde{\kappa}^2C}{4R^3} \ ,
\end{equation}
which can be recast as
\begin{eqnarray} \label{eq:Rtautau2}
R_{\tau\tau}&=&\frac{R_\tau^2}{R}+\frac{1}{R}\left(1-\frac{3M}{R}+\frac{2Q^2}{R^2}\right) \nonumber \\
&&+\frac{\gamma}{R^3} {\sqrt{1-\frac{2M}{R}+\frac{Q^2}{R^2}+R_\tau^2}}\ .
\end{eqnarray}

We can then follow the same procedure as in the Schwarzschild case. It is convenient, though, to use dimensionless variables of the form
 \begin{eqnarray}
R&\to & M x \nonumber \\
\tau &\to & M t \nonumber \\
Q^2 &\to & M^2 y \nonumber\\
\gamma &\to & M^2\tilde{\gamma}  \nonumber
\end{eqnarray}

\begin{figure}[t!]
\includegraphics[width=0.45\textwidth]{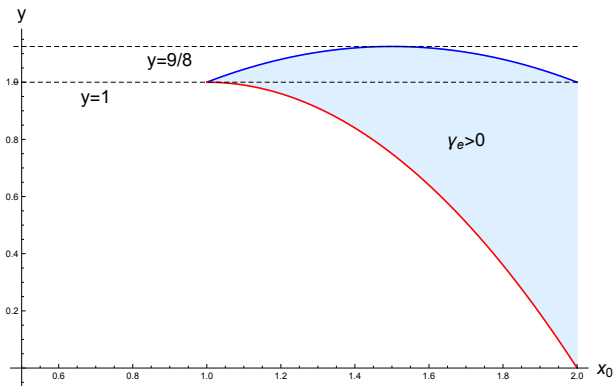}
\caption{The blue region enclosed between the curves $y=x_0(3-x_0)/2$ and $y=x_0(2-x_0)$   represents the allowed values of $x_0\equiv R_0/M$ as a function of the squared charge-to-mass ratio $y\equiv Q^2/M^2$ for which equilibrium configurations with positive energy density ($\tilde{\gamma}_e>0$) are possible in thin-shell wormholes of surgically joined Reissner-Nordstr\"om space-times. For $x_0>2$ the band $0<y<1$ always has positive energy density.}\label{fig:PositiveEnergy}
\end{figure}

In terms of these variables the horizons of the Reissner-Nordstr\"om configurations are located at $x_{\pm}=1\pm \sqrt{1-y^2}$ if the (squared) charge-to-mass ratio satisfies $0<y<1$, while no horizons are present when $y>1$ (with $y=1$ corresponding to extremal configurations). On the other hand, using these variables the differential equation (\ref{eq:Rtautau2}) becomes
\begin{eqnarray}\label{eq:xtt}
x_{tt}&= &\frac{x_t^2}{x}+\frac{1}{x}\left(1-\frac{3}{x}+\frac{2y}{x^2}\right)
\nonumber \\
&&+\frac{\tilde{\gamma}}{x^3} \sqrt{1-\frac{2}{x}+\frac{y}{x^2}+x_t^2}\,.
\end{eqnarray}
The equilibrium points follow by imposing $x=x_0$ (setting the throat) with $x_t=x_{tt}=0$, which requires to tune the free parameter $\tilde{\gamma}$ to the particular value
\begin{equation}
\tilde{\gamma}_e=\frac{x_0(3 x_0-x_0^2-2 y)}{ \sqrt{{x_0^2-2 x_0+y}}} \,,
\end{equation}
which is the natural extension of (\ref{eq:gammae}) for the Reissner-Nordstr\"om configurations. This condition has a clear impact on the kind of matter source able to produce equilibrium configurations. In fact, from the definition $\tilde{\gamma} \equiv \frac{3\tilde{\kappa}^2 C}{4M^2}$, it is immediate  to see that the sign of $\tilde{\gamma}_e$ determines the sign of the energy density of the thin shell, as follows from Eq.(\ref{eq:sigmacon}). The possibility of having positive energy density configurations is thus not ruled out {\it a priori}. Other equilibrium configurations correspond to the point $(x_0,y)=(1,1)$ for any $\tilde{\gamma}_e$, and the curve $y=x_0(3-x_0)/2$ for $\tilde{\gamma}=0$ (see the blue curve in Fig.\ref{fig:PositiveEnergy}).

Analyzing this expression, we see that for $0<y<1$ we have $\tilde{\gamma}_e>0$ if $x_0>1+\sqrt{1-y}$ (equivalently $y>1-(x_0-1)^2$), i.e., only when the shell is located outside the would-be external horizon. This means, in particular, that in the Schwarzschild limit, $y\to 0$, we must have $x_0>2$, whereas in the extremal limit $y\to 1$ we must have $x_0>1$. Beyond extremality, we also have $\tilde{\gamma}_e>0$ when $1<y<9/8$ if the radius of the shell is located within the interval
\begin{equation}
\frac{3-\sqrt{9-8y}}{2}<x_0<\frac{3+\sqrt{9-8y}}{2}
\end{equation}
or, equivalently, $x_0(2-x_0)<y<x_0(3-x_0)/2$, as shown in Fig.\ref{fig:PositiveEnergy}. Note that the parabola $y=x_0(3-x_0)/2$ represents equilibrium points with $\tilde{\gamma}_e=0$ and that the point $(x_0,y)=(1,1)$ is also an equilibrium point for arbitrary $\tilde{\gamma}_e$.

\begin{figure}[t!]
\includegraphics[width=0.45\textwidth]{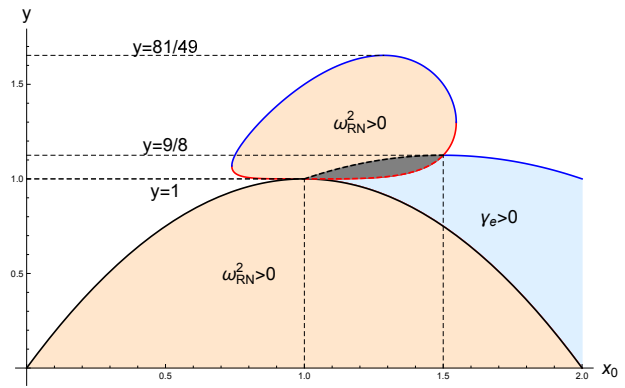}
\caption{Representation of the domain where $\omega_{RN}^2>0$ in the Reissner-Nordstr\"om scenario outside its horizons (as given by the condition (\ref{eq:domain2})) and its overlap (gray shadowed region) with $\tilde{\gamma}_e>0$. The region enclosed by the big parabola with vertex at $(1,1)$ (given by Eq.(\ref{eq:domain0})) is contained within the horizons and, therefore, is not of physical interest.}\label{fig:overlap}
\end{figure}

\begin{figure}[t!]
\includegraphics[width=0.40\textwidth]{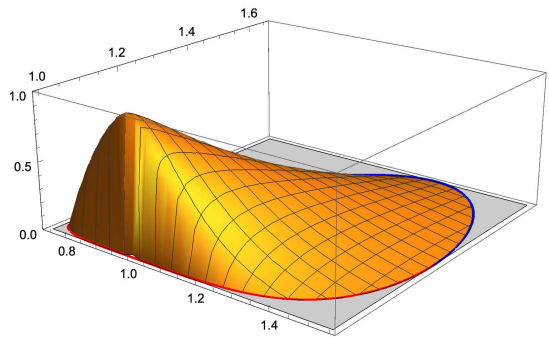}
\caption{Representation of $\omega_{RN}^2>0$ in the egg-shaped domain (\ref{eq:domain2}) shown in Fig.\ref{fig:overlap}. }\label{fig:domain2a}
\end{figure}

Let us now focus on the stability of the solutions. Expanding the field equation \eqref{eq:xtt} around an equilibrium point $x_0$, to linear order in $\delta(t) \equiv x(t)-x_0$ one finds
\begin{equation} \label{eq:RHSeqRN}
\delta_{tt} +\omega_{RN}^2 \delta(\tau) =0
\end{equation}
so that the stability condition translates into
\begin{eqnarray} \label{eq:stabilityRN1}
&\omega_{RN}^2=- \frac{\left[(2 (x_0-4) x_0+9) x_0^2+(3 x_0-8) x_0 y+2 y^2\right]}{x_0^4 \left[(x_0-2) x_0+y\right]} >0 \,.
%\nonumber
\end{eqnarray}
It turns out that this condition can only be satisfied either  i) when $0<y<1$ if
\begin{equation} \label{eq:domain0}
1-\sqrt{1-y}<x_0<1+\sqrt{y-1}
\end{equation}
(equivalently, $y<x_0(2-x_0)$), which corresponds to the region between the two horizons, or ii) when $1<y<81/49$ if
\begin{equation}\label{eq:domain2}
y_-(x_0)<y<y_+(x_0) ,
\end{equation}
where we have defined the two curves
\begin{equation}
y_\pm(x_0)=\frac{x_0}{4}\left(8-3x_0\pm\sqrt{7(x_+-x_0)(x_0-x_-)}\right)
\end{equation}
and  $x_\pm=2(4\pm\sqrt{2})/7$. These two regions are plotted in Fig.\ref{fig:overlap} (filled in orange) together with the domain corresponding to positive energy density configurations (filled in blue). As it can be seen there, there is a non-zero subset of overlapping points (filled in gray) such that
\begin{equation} \label{eq:photo}
1<y<9/8 \ ,
\end{equation}
and bounded by the curves
\begin{equation} \label{eq:boundcurv}
y_-(x_0)<y<\frac{x_0}{2}\left(3-x_0\right) \ ,
\end{equation}
and resembling an egg-shaped domain (see Fig.\ref{fig:domain2a}). It should be pointed out that the condition (\ref{eq:photo}) exactly coincides with the condition for having a photon sphere in the full geometry of the overcharged Reissner-Nordstr\"om solution. The existence of this overlapping region puts forward that, unlike in GR, within the $f(\mathcal{R})$ Palatini framework it is possible to have stable, traversable thin-shell wormhole solutions supported by sources with positive energy density. Though the negative energy density region is much larger than the positive one, it is remarkable that stable solutions of this type (satisfying all the energy conditions regardless of the form of the $f(\mathcal{R})$ Lagrangian) can exist.

\begin{figure}[t!]
\includegraphics[width=0.40\textwidth]{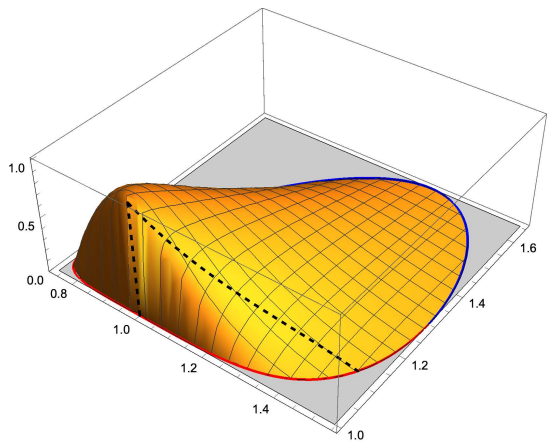}
\caption{A different view of the surface $\omega_{RN}^2>0$ in the domain (\ref{eq:domain2}) shown in Fig.\ref{fig:overlap} including the projection of the curve $y=\frac{x_0}{2}\left(3-x_0\right)$ (dashed black curve)  on this surface to highlight the region with positive energy density.}\label{fig:domain2b}
\end{figure}

Particular attention should be paid to the parabolic segment $y=x_0(3-x_0)/2$ between $x_0=1$ and $x_0=3/2$ (plotted in Fig.\ref{fig:domain2b} as a black dashed curve) that represents the upper bound of the overlapping region because it corresponds to the curve of equilibrium points with $\tilde{\gamma}=0$, which have vanishing energy density. Our analysis indicates that such a curve must represent stable equilibrium points.

%%%%%%%%%%%%%%%%%%%%%%%%%%%%%%%%%%%%%%%%%%%%%%%%%%%%%%%%%%%%%%%%%
\section{Numerical analysis} \label{sec:IV}
%%%%%%%%%%%%%%%%%%%%%%%%%%%%%%%%%%%%%%%%%%%%%%%%%%%%%%%%%%%%%%%%%

\begin{figure}[t!]
\includegraphics[width=0.40\textwidth]{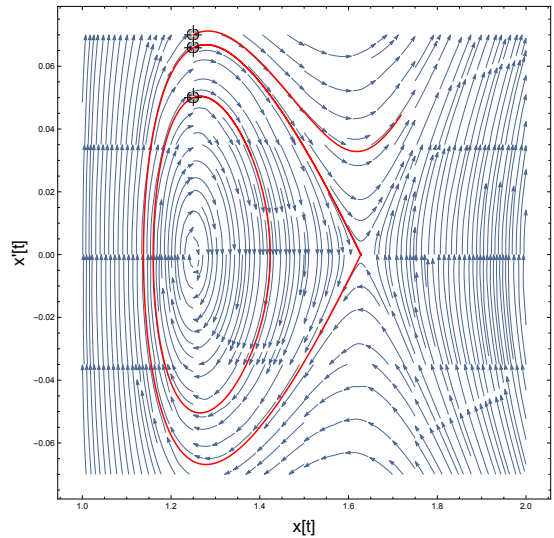}
\caption{Phase portrait of positive energy configurations located at $x_0=5/4$ with $y=1.05$. The red curves represent numerical integrations with initial velocity $\dot x_0=0.07$ (unstable solution) $\dot x_0=0.05$ (stable inner solution), and $\dot x_0 \approx 0.06586$ (limiting stable solution). A locator is placed at the initial point of each integration. }\label{fig:PEPP}
\end{figure}

\begin{figure}[t!]
\includegraphics[width=0.4\textwidth]{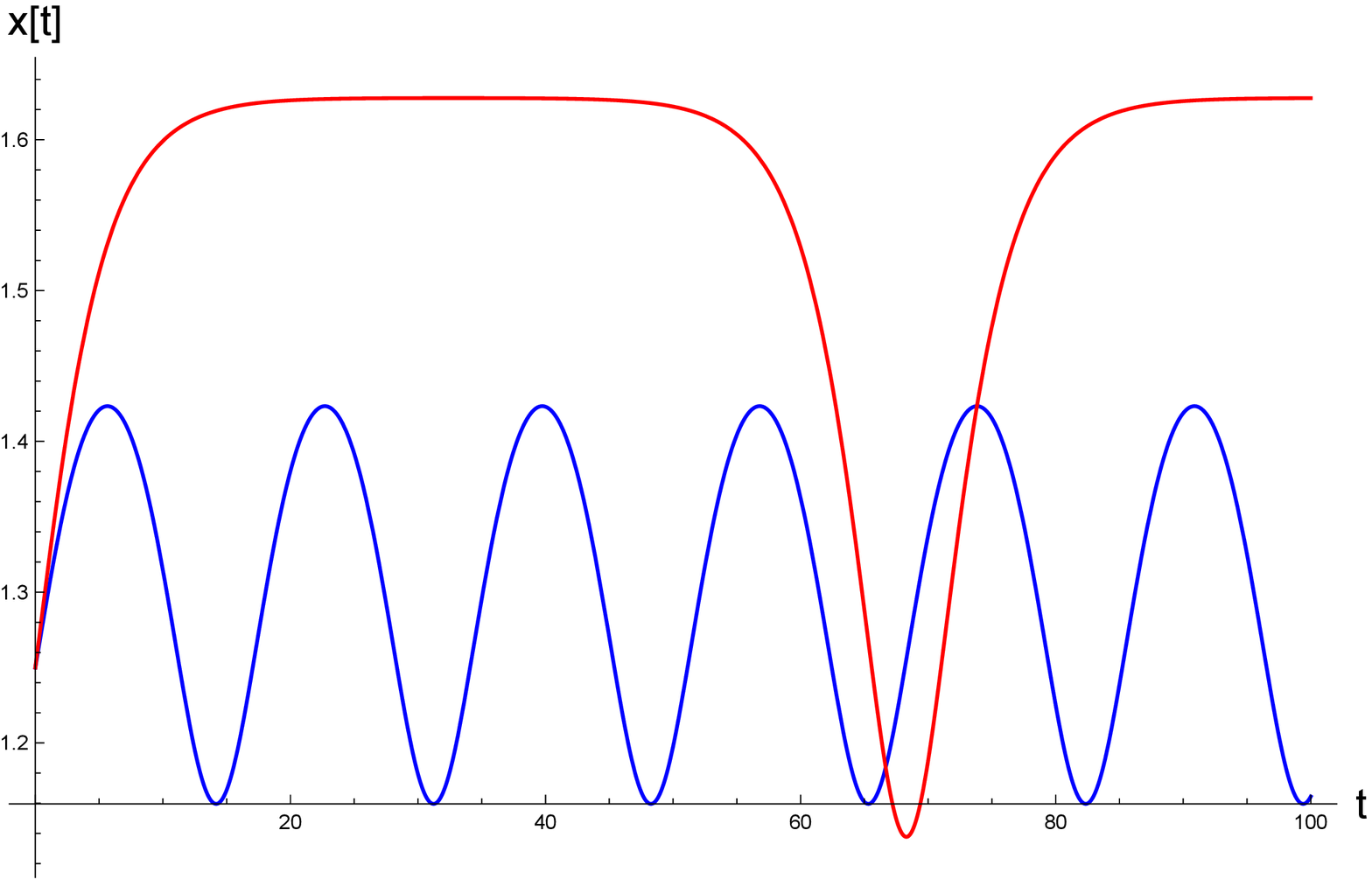}
\caption{Time evolution of $x(t)$ around $x_0=5/4$  with $y=1.05$ for the two stable solutions of Fig.\ref{fig:PEPP}. The case $\dot x_0=0.05$ (blue) has a higher oscillation frequency. As one approaches the limiting stable solution the period grows and the maximum amplitude flattens.   }\label{fig:Postable}
\end{figure}

\begin{figure}[t!]
\includegraphics[width=0.4\textwidth]{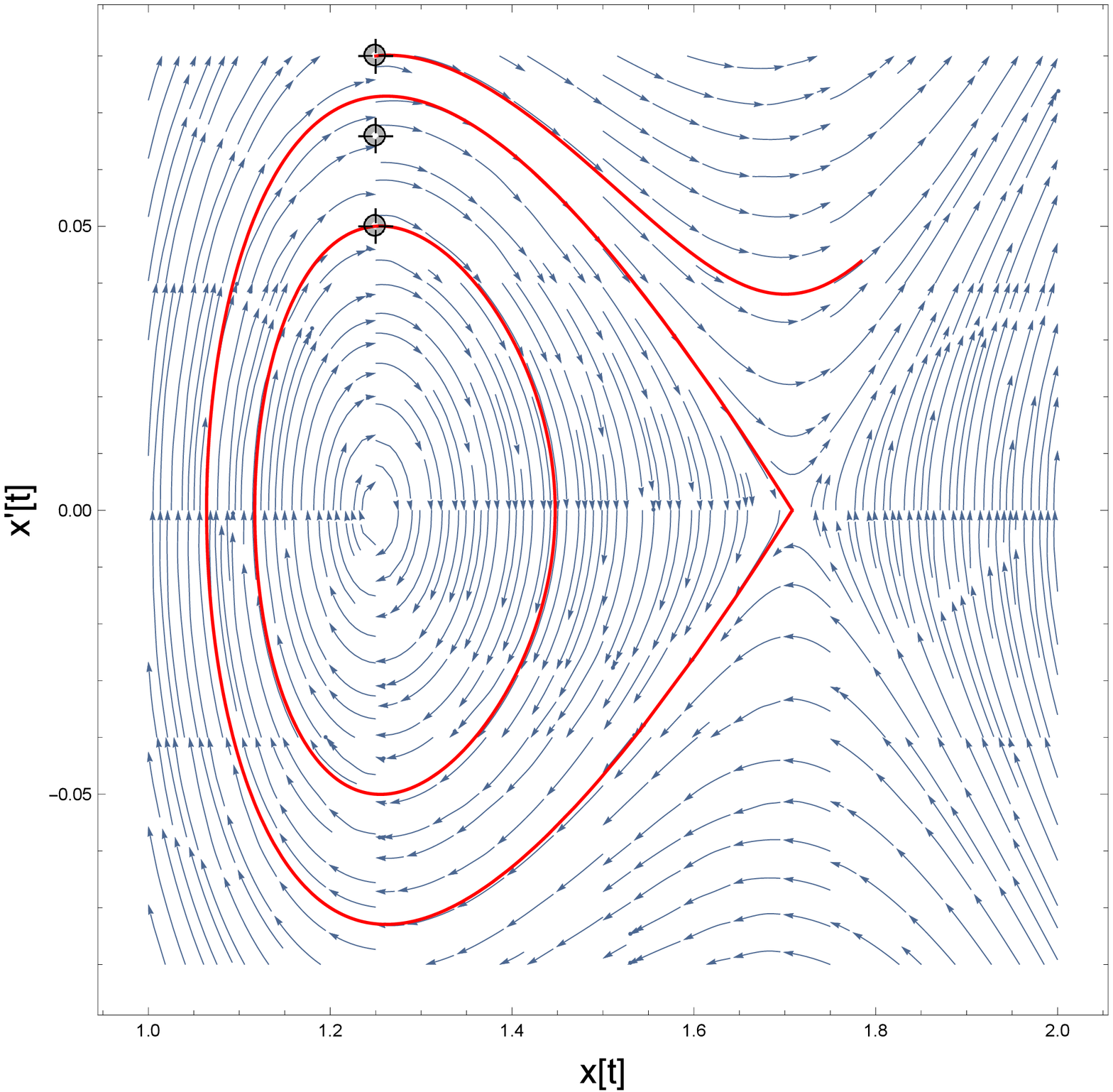}
\caption{Phase portrait of negative energy configurations located at $x_0=5/4$ with $y=1.5$. The red curves represent numerical integrations with initial velocity $\dot x_0=0.08$ (unstable solution) $\dot x_0=0.05$ (stable inner solution), and $\dot x_0\approx 0.07282$ (limiting stable solution). A locator is placed at the initial point of each integration. }\label{fig:NEPP}
\end{figure}

\begin{figure}[t!]
\includegraphics[width=0.4\textwidth]{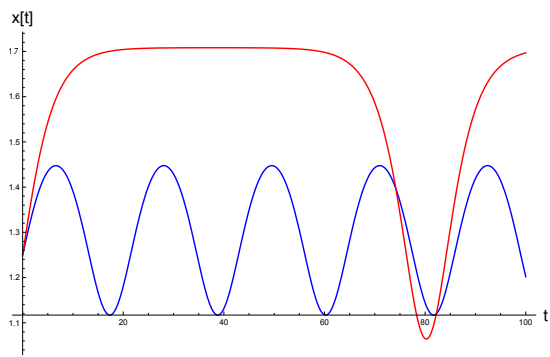}
\caption{Time evolution of $x(t)$ around $x_0=5/4$ with $y=1.5$ for the two stable solutions of Fig.\ref{fig:NEPP}. The case $\dot x_0=0.05$ (blue) has a higher oscillation frequency. As one approaches the limiting stable solution the period grows and the maximum amplitude flattens much like in the positive energy case.}\label{fig:Negstable}
\end{figure}

We will now present some numerical results for the various possible thin-shell wormhole configurations of the Reissner-Nordstr\"om type studied before, where stable equilibrium points may arise. From our previous analysis, we concluded that one can find stable configurations with positive, negative, and zero energy density. We also identified a peculiar point at $(x_0,y)=(1,1)$, an extremal configuration, which deserves special attention. For simplicity and generality, we will present portrait diagrams of the system for different choices of the parameters $x_0$ and $y$ to highlight the regions of stability.

\begin{figure}[t!]
\includegraphics[width=0.4\textwidth]{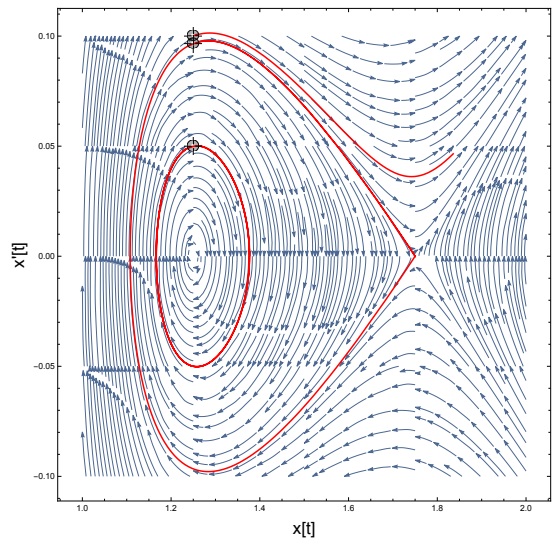}
\caption{Phase portrait of zero energy configurations located at $x_0=5/4$ with $y=35/32\approx 1.09375$. The red curves represent numerical integrations with initial velocity $\dot x_0=0.1$ (unstable solution) $\dot x_0=0.05$ (stable inner solution), and $\dot x_0 \approx 0.09658$ (limiting stable solution). A locator is placed at the initial point of each integration. }\label{fig:ZEPP}
\end{figure}

\begin{figure}[t!]
\includegraphics[width=0.4\textwidth]{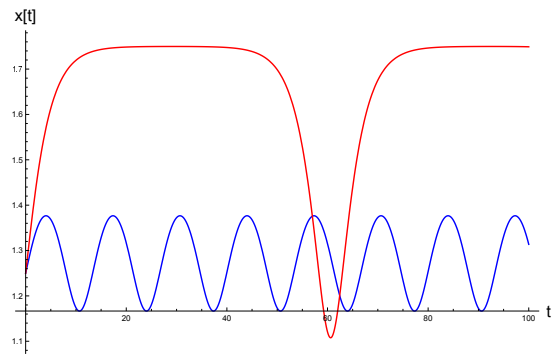}
\caption{Time evolution of $x(t)$ around $x_0=5/4$ and $y=35/32\approx 1.09375$ for the two stable solutions. The qualitative behavior is very similar to the positive and negative energy cases.  }\label{fig:Zerostable}
\end{figure}

\begin{figure}[t!]
\includegraphics[width=0.48\textwidth]{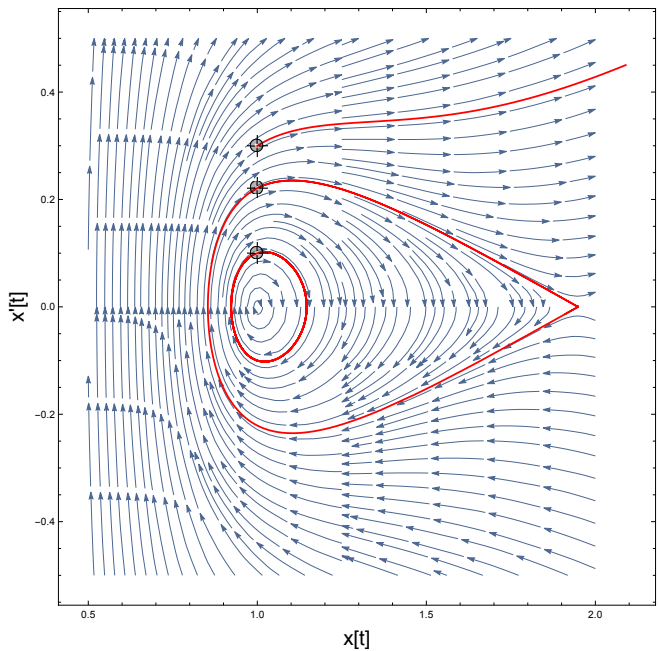}
\caption{Phase portrait of the extremal configuration with $x=1$ and $y=1$. The red curves represent numerical integrations with initial velocity $\dot x_0=0.33$ (unstable solution) $\dot x_0=0.1$ (stable inner solution), and $\dot x_0 \approx 0.02209$ (limiting stable solution). A locator is placed at the initial point of each integration. }\label{fig:XtPP}
\end{figure}

\begin{figure}[t!]
\includegraphics[width=0.4\textwidth]{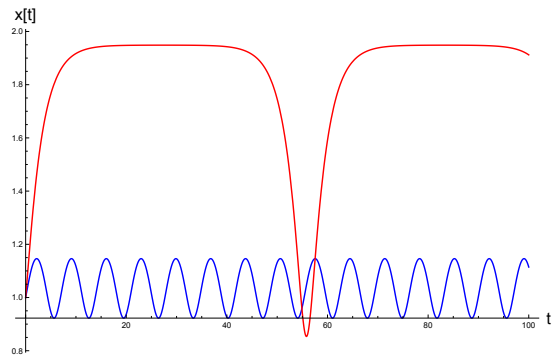}
\caption{Time evolution of $x(t)$ around $x_0=1$ with $y=1$ for the two stable solutions of Fig.\ref{fig:XtPP}. Though the qualitative features are similar to the previous cases, the stability domain in bigger and the amplitude of the limiting stable configuration is also larger. }\label{fig:Xtstable}
\end{figure}

Positive energy stable solutions are shown in Figs.\ref{fig:PEPP} and \ref{fig:Postable}, negative energy solutions in Figs.\ref{fig:NEPP} and \ref{fig:Negstable}, and zero energy solutions in Figs.\ref{fig:ZEPP} and \ref{fig:Zerostable}. The extremal case is presented in  Figs.\ref{fig:XtPP} and \ref{fig:Xtstable}. As one can see, in all four cases the qualitative features of the various plots are similar. Nonetheless, it is important to note that the domain of stability in the extremal case is much larger than in the overextremal configurations. The amplitude of the oscillations of the limiting stable curve is also larger in the extremal case, being smaller in the positive energy configuration.

%%%%%%%%%%%%%%%%%%%%%%%%%%%%%%%%%%%%%%%%%%%%%%%%%%%%%%%%%%%%%%%%%
\section{Conclusion} \label{sec:V}
%%%%%%%%%%%%%%%%%%%%%%%%%%%%%%%%%%%%%%%%%%%%%%%%%%%%%%%%%%%%%%%%%

In this work we have employed the recently developed formalism of the junction conditions for Palatini $f(\mathcal{R})$ theories of gravity in order to study spherically symmetric traversable thin-shell wormholes. We have found the relations between curvature and matter fields at the shell, whose main highlight is the fact that the effective number of degrees of freedom is reduced from two (which are the ones found in GR and in metric $f(R)$ gravity) down to one. This scenario actually describes the dynamics of massless stress-energy fields on the shell, something which cannot be found in GR. Moreover, in the simplest possible scenario, namely, the one  where the (normal) discontinuity of the stress-energy tensor across the shell is vanishing, the energy density can be resolved as an inverse cubic power of the radius of the shell, without any {\it a priori} restrictions on its sign and, therefore, on the fulfillment of the energy conditions.

Subsequently we have used this result to build traversable thin-shell wormholes from surgically grafted spherically symmetric vacuum space-times. We have shown that, no matter the shape of the Palatini $f(\mathcal{R})$ theory chosen (except for the trivial choice $f(\mathcal{R})=\mathcal{R}-2\Lambda$, corresponding to the Einstein-Hilbert action with a cosmological constant term), such wormholes can arise from two vacuum space-times joined at a certain shell radius beyond its Schwarzschild radius (in order to avoid the presence of horizons). However, the thin-shell wormholes built this way have a growing mode under linear perturbations no matter the shape of the $f(\mathcal{R})$ chosen, which renders the corresponding solutions linearly unstable. Thus we turned our attention to  thin-shell wormholes from surgically joined Reissner-Nordstr\"om space-times (supported by electrovacuum fields), studying the regions with positive energy density, $\tilde{\gamma}_e>0$, and, separately, the regions with stable configurations, $\omega_{RN}^2>0$. It turns out that there exists a non-vanishing overlapping region where both conditions can be met at the same time, as given by a constraint on the (squared) mass-to-charged ratio, $1<y<9/8$ (corresponding to the overcharged Reissner-Nordstr\"om solution having a photon sphere), and on the radius of the shell bounded by certain curves, as defined by Eq.(\ref{eq:boundcurv}).  Therefore, traversable thin-shell wormholes supported by matter sources fulfilling the energy conditions are actually possible within these surgically joined Reissner-Nordstr\"om geometries of Palatini $f(\mathcal{R})$ gravity. Remarkably, we also find a subset of stable solutions which lies between the positive energy and negative energy solutions, thus representing thin shells with vanishing energy density. Such configurations can thus only be supported by the topological electric flux and are a peculiar feature of the Palatini $f(\mathcal{R})$ framework.

It is worth mentioning that, though in this work we only considered symmetric thin-shell wormhole space-times (which means that the solutions on $\mathcal{M}_{\pm}$ are the same), nothing in our construction prevents us to generalize them to two distinct surgically joined space-times, as long as Eq.(\ref{eq:flux}) is satisfied. Such a kind of asymmetric thin-shell wormholes have been recently explored in the literature  \cite{Forghani:2018gza}. Interestingly, in such solutions a novel shadow may emerge, related to the existence of different photon spheres on each side, such that an observer on the side of the wormhole with lower mass would see, in addition to the images of its own photon sphere, also those photons bouncing back from the photon sphere on the other side across the shell \cite{Wang:2020emr,Wielgus:2020uqz}. The observations of such double shadows could be a hint on the existence of new physics beyond the canonical compact objects of GR.

To conclude, the results obtained in this paper proves that the modifications in the junction conditions when considering   $f(\mathcal{R})$ gravity in the Palatini formulation yield non-trivial novelties in the analysis of traversable thin-shell wormholes as compared to their GR or metric $f(\mathcal{R})$ counterparts. Moreover, this puts forward the new possibilities offered by these theories in any scenarios requiring a matching between two regions, such as in cosmic strings, domain walls, stellar models, or within the membrane paradigm \cite{Maggio:2020jml}, topics which are of both astrophysical and cosmological interest. We hope to further report on these issues soon.

%%%%%%%%%%%%%%%%%%%%%%%%%%%%%%%%%%%%%%%%%%%%%%%%%%%%%%%%%%%%%%%%%
\section*{Acknowledgements}
%%%%%%%%%%%%%%%%%%%%%%%%%%%%%%%%%%%%%%%%%%%%%%%%%%%%%%%%%%%%%%%%%

FSNL acknowledges support from the Funda\c{c}\~{a}o para a Ci\^{e}ncia e a Tecnologia (FCT) Scientific Employment Stimulus contract with reference CEECIND/04057/2017, and thanks the FCT research projects No. UID/FIS/04434/2020 and No. CERN/FIS-PAR/0037/2019. GJO is funded by the Ramon y Cajal contract RYC-2013-13019 (Spain). DRG is funded by the \emph{Atracci\'on de Talento Investigador} programme of the Comunidad de Madrid (Spain) No. 2018-T1/TIC-10431, and acknowledges further support from the Ministerio de Ciencia, Innovaci\'on y Universidades (Spain) project No. PID2019-108485GB-I00/AEI/10.13039/501100011033, and the FCT project No. PTDC/FIS-PAR/31938/2017.
This work is supported by the Spanish project  FIS2017-84440-C2-1-P (MINECO/FEDER, EU), the project H2020-MSCA-RISE-2017 Grant FunFiCO-777740, the project SEJI/2017/042 (Generalitat Valenciana), the Consolider Program CPANPHY-1205388, the Severo Ochoa grant SEV-2014-0398 (Spain), the FCT project PTDC/FIS-OUT/29048/2017,  and the Edital 006/2018 PRONEX (FAPESQ-PB/CNPQ, Brazil, Grant 0015/2019). This article is based upon work from COST Actions CA15117  and CA18108, supported by COST (European Cooperation in Science and Technology).

%%%%%%%%%%%%%%%%%%%%%%%%%%%%%%%%%%%%%%%%%%%%%%%%%%%%%%%%%%%%%%%%%

%%%%%%%%%%%%%%%%%%%%%%%%%%%%%%%%%%%%%%%%%%%%%%%%%%%%%%%%%%%%%%%%%

%%%%%%%%%%%%%%%%%%%%%%%%%%%%%%%%%%%%%%%%%%%%%%%%%%%%%%%%%%%%%%%%%
\end{document}